\documentstyle[aps,preprint]{revtex}

\tightenlines

\begin{document}

\title{Electron-phonon interaction in C$_{70}$}

\author{D. Provasi$^{1}$, N. Breda$^{1,2}$, R.A. Broglia$^{1,3,4}$, 
	G. Col\`o$^{1,3}$, H.E. Roman$^{3}$ and G. Onida$^{5}$}
	
\address{$^1$Dipartimento di Fisica, Universit\`a di Milano,
	 Via Celoria 16, I-20133 Milano, Italy}

\address{$^2$INFM, Sezione di Milano, Via Celoria 16, I-20133 Milano, Italy}

\address{$^3$INFN, Sezione di Milano, Via Celoria 16, I-20133 Milano, Italy}

\address{$^4$The Niels Bohr Institute, University of Copenhagen,
	 D-2100 Copenhagen, Denmark}

\address{$^5$Istituto Nazionale per la Fisica della Materia -
	 Dipartimento di Fisica dell' Universit\`a di Roma Tor Vergata,
	 Via della Ricerca Scientifica, I--00133 Roma, Italy}

\date{\today}
\maketitle

\begin{abstract}
The matrix elements of the deformation potential of C$_{70}$ are calculated
by means of a simple, yet accurate solution of the electron-phonon
coupling problem in fullerenes, based on a parametrization of the
ground state electronic density of the system in terms of $sp^{2+x}$ 
hybridized orbitals.
The value of the calculated dimensionless 
total electron-phonon coupling
constant is $\lambda\approx0.1$, an order of magnitude smaller than in 
C$_{60}$, consistent with the lack of a superconducting
phase transition in C$_{70}$A$_3$ fullerite, and in overall
agreement with measurements of the broadening of Raman peaks in C$_{70}$K$_4$. 
We also calculate the photoemission cross section of C$_{70}^-$, which is
found to display less structure than that associated with C$_{60}^-$, in 
overall agreement with the experimental findings.   
\end{abstract}

\bigskip
\leftline{PACS: 74.70.Wz, 63.20.Kr, 33.60.-q, 61.48.+c}

\newpage
\section{Introduction}
The discovery that some alkali-doped C$_{60}$ compounds are superconducting
\cite{Flemin,Hebard,Hokzer,Rossem,Taniga} with a transition temperature as
high as 40 K for the fulleride C$_{60}$Cs$_3$ under pressure 
\cite{Palstr} has triggered considerable interest in the study of
fulleride doped materials at low temperatures. Much effort has been
concentrated in trying to understand whether such large values of $T_c$ can
be caused by the coupling of electrons to phonons. The electron-phonon
coupling constants for C$_{60}$ have been estimated both from experiments 
\cite{Winter} and calculations \cite{Gunnar}. There exists some consensus on 
the fact that C$_{60}$ compounds are
$s-$wave BCS-like superconductors driven by the coupling to selected
intramolecular phonons \cite{Gunnar}. On the other hand, no calculation of this 
coupling exists in the literature for the case of C$_{70}$. 
Since the associated alkali compounds display no superconductivity, 
with a smooth and continuous behavior of the resistivity down to 
about 1 K~\cite{Wang}, it is an important issue 
to compare the electron-phonon coupling constant for C$_{70}$ with those of 
C$_{60}$.

In the present paper we address this problem from a theoretical point of view,
computing the deformation potential of C$_{70}$ and
extracting the associated total electron--phonon coupling constant, $\lambda$.
It will be concluded that the resulting value of $\lambda$ is much smaller 
than in C$_{60}$, essentially ruling out the possibility of a superconducting
phase at low temperatures for doped fullerides built out of C$_{70}$. 

This negative result, which can be understood in rather general, qualitative 
structural terms, as well discussed at the end of Sec. III, has two positive
oucomes: first, it provides further evidence for the $s-$wave BCS-like 
mechanism at the basis of alkali-doped fullerides superconductivity. Second, it
sheds light on the properties fullerenes have to exhibit to be ``good'' building 
blocks of eventual organic superconductors, indicating the direction of the 
highly symmetric, small fullerenes (cf. also Refs.~\cite{Cot98,Dev98}), a 
strategy which seems to be universally valid also for other compact systems, 
like the atomic nucleus~\cite{Bar_tbp}.  

In the production of fullerene clusters by means of laser vaporization or arc
discharge the yield of C$_{70}$ can be optimized, by varying the experimental 
conditions~\cite{Sma88,Kra90}, up to a ratio of about 10\% of C$_{70}$ with 
respect to  C$_{60}$. This allows the production of sufficiently large amounts
of C$_{70}$ for the synthesis of C$_{70}$ molecular crystals~\cite{Tea93}.
Hence, the structural and electronic properties of C$_{70}$ can be 
experimentally studied,
and compared with those of C$_{60}$. In analogy with the case of solid 
C$_{60}$, C$_{70}$ crystals can be doped with alkali metals, and the 
conductivity of the resulting compounds can be investigated.

On the theoretical
side, {\em ab-initio} methods based on the local density approximation
(LDA)~\cite{Wand92} have been applied to the study of the 
ground-state properties of C$_{70}$.
The ionic configuration, characterized by a D$_{5h}$ geometry, is
consistent with the NMR measurements~\cite{NMR}, and the electronic 
structure agrees
reasonably well with UPS data. 
We adopt in this work the ground-state geometry of
Ref.~\cite{Wand92}. The vibrational properties of C$_{70}$
have been studied in~\cite{LDA_phon},
starting with the same geometry. 
A simplified
model like the bond charge model (BCM) is nevertheless 
able to reproduce the experimental values for the optically 
active frequencies within 4\%~\cite{Benede}. Indeed, beyond the
electronic and vibrational properties, {\it ab--initio} methods also 
allow to compute the coupling between these two degrees
of freedom without the need of introducing any adjustable parameter.
However, {\it ab-initio} methods in the case of large clusters,
such as C$_{60}$ and C$_{70}$, are
computationally demanding and may not be particularly transparent. 
As an alternative, in order 
to extract the electron-phonon coupling constants, it is possible to
adopt a simplified
numerical method which already proved to be satisfactory in the case of 
C$_{60}$~\cite{BredaN}. This method, albeit approximated, has been shown to 
lead to quite satisfactory results, also because it is 
devised to take into account accurate phonon eigenvectors, which are known to influence very strongly
the matrix elements of the deformation potential. In fact, the relative 
strength of the latter is strictly related to the bond stretching 
associated to the ionic displacements.
In the present work, using the scheme 
introduced in ref.\cite{BredaN} for  C$_{60}$, we address the case of C$_{70}$.

\section{Electron-phonon coupling}

Our calculation starts with the determination of the 
electronic states of fullerene-C$_{70}$, and the corresponding electron 
density, within DFT-LDA.
We include the 
exchange and correlation effects according to the parametrization of
Perdew and Zunger~\cite{Perdew} of the Monte-Carlo results of Ceperley
and Alder~\cite{Cepal}. 
No GGA corrections are considered, since LDA is known to yield very accurate 
results for the equilibrium geometry 
and charge density in such covalently bonded carbon sistems~\cite{artGGA}.
The role of carbon atoms is taken
into account using {\it ab-initio} norm-conserving pseudopotentials 
\cite{Bachel}. A spherical basis, made up with states which are solution of 
a central potential which mimics the $L$=0 component of the LDA local
potential, is employed. Spherical wavefunctions with $n$ and $l$
respectively up to 15 and 20 are included in the basis 
(this gives $\approx$ 4000 basis vectors). 
The Kohn-Sham equations are solved in matrix form on this basis,
as described in more detail in Ref.~\cite{Ala95}. 
In that work the maximum values of  $n$ and $l$ were slightly different,
but the results are essentially the same. 
The HOMO-LUMO gap is 1.87 eV, in agreement with other LDA 
calculations~\cite{Wand92}. As in C$_{60}$, errors due to the neglect
of both self-energy and excitonic effects nearly cancel each other, leading to
a theoretical value close to the experimentally measured optical gap 
(1.6$\pm$0.2 eV). The Kohn-Sham levels up to about -10 eV are in overall
agreement with the UPS spectra. The symmetries of the HOMO and LUMO 
levels ($E_1^{\prime\prime}$) are the same
as those obtained in~\cite{Wand92}. The different results of Ref.~\cite{Sai91}
are probably an artifact due to the limited basis set used there.  

The energies and eigenvectors of the vibrational normal modes have been taken 
from a BCM calculation ~\cite{BCM}. Then, 
following~\cite{BredaN}, we have parametrized
the {\it ab-initio} electronic density in terms of $sp^{2+x}$ hybrid 
orbitals obtained as linear combinations of the four $s$ and $p$ valence orbitals of
each carbon atom.
Three of the hybrid orbitals are directed along the bonds connecting
the atom with its three nearest neighbors, while the fourth is determined by 
orthogonality conditions and takes care of the additional $\pi$-bonding 
present in the fullerenic cages. 
The radial wave functions of the $s$ and 
$p$ orbitals are taken as standard Slater wave functions ($R_{s} = 
\frac{2}{\sqrt{\sigma_{1}^{3}}}e^{(-r/\sigma_{1})}$ and 
$R_{p} = \frac{2}{\sqrt{3\sigma_{2}^{5}}}re^{(-r/\sigma_{2})}$), where the  
parameters $\sigma_{1}$ and $\sigma_{2}$ have been adjusted in order to 
obtain the best fit to the  C$_{70}$ charge density computed in LDA. This 
leads to $\sigma_{1}=0.66\ \AA$ and $\sigma_{2}=0.36\ \AA$ (to be 
compared with similar values
obtained in the case of C$_{60}$~\cite{BredaN}). For comparison,
the atomic values are $\sigma_{1}=0.65\ \AA$ and 
$\sigma_{2}=0.17\ \AA$.

After having determined the hybrid orbitals, we can write the contribution to 
the total electronic density arising from a single atom, and carry out a 
multipole expansion of it around the center of the cluster. 
Adding the contributions of the 70 atoms one obtains the total density. 
The deformation potential associated with a normal mode $\alpha$ is then
\begin{equation}
 \left. V_{def}^{(\alpha)} (\vec r)  =  \sum_{N=1}^{70} \vec Q_{N}^{(\alpha)}
 \cdot \vec \nabla_{N} V_{TOT}\ (\vec r, \{\vec R\}) \right|_{\{\vec R\} 
 = \{\vec {R^0}\}},
\label{defpot}\end{equation}
where $\vec Q_{N}^{(\alpha)}$ is the displacement of the $N$-th ion, 
$V_{TOT}$ is the total LDA electronic potential, and $\{ \vec R\}$ represents 
the whole set of ionic coordinates whose equilibrium values are 
$\{ \vec R^0\}$ ($\vec Q_N = \vec R_N - \vec R_N^0$). The pseudopotential 
term of $V_{TOT}$ displays an explicit dependence on the ionic
positions, therefore the calculation of its gradient presents no 
difficulties. On the other hand, the Hartree and exchange-correlation (XC)
terms depend on $\{\vec R\}$ 
{\em implicitly} through the electronic density $\varrho$. We thus write 
\begin{equation}
 \sum_{N} \vec Q_N \cdot \vec \nabla_N\ V_{\rm Hartree,XC}
 [\varrho(\vec r,\{\vec R\})] = \sum_{N} \frac{\partial V_{\rm Hartree,XC}
 [\varrho]}{\partial \varrho} \vec Q_N \cdot \vec \nabla_N\  
 \varrho(\vec r,\{\vec R\}).
\label{dercomp}\end{equation}
One can easily calculate $\partial V/\partial \varrho$, while the gradient 
of the total density is determined by assuming, within the present hybrid 
orbital model, that upon moving the ions according to the 
phonon eigenvectors  the hybrid orbitals change their direction but 
{\em not}  
their shape. This makes the calculation quite simple and transparent, 
since it is possible to 
visualize the effect of 
a normal ionic displacement on the electronic orbitals. Finally, we can 
evaluate the matrix elements of the deformation potential, 
$[V_{def}^{(\alpha)}]_{ij} \equiv \langle i | V_{def}^{(\alpha)} | j \rangle$, 
between Kohn-Sham wavefunctions. 

If we are interested in the behavior of the conductivity in alkali-doped 
fullerites, or in the photoemission spectrum of C$_{70}$, we should start
from phonon energies, and deformation potentials, evaluated by using the 
electronic density of a negative charged ion. On the other hand, because the
density of the 280 valence electrons of C$_{70}$ is not appreciably altered 
by adding a few more electrons, one expects the matrix elements of the 
deformation potentials associated with C$_{70}$ and with C$_{70}^{n-}$ 
($1\le n\le 4$), to be rather similar. 

We evaluate, for each phonon
$\alpha$, the {\em reduced} matrix elements $[V_{def}^{(\alpha)}]_{ii}$ 
(corresponding to a given electronic state $i$) which are 
denoted by $g_\alpha$, as 
well as the partial electron-phonon coupling constants 
$\lambda_\alpha$ defined as
\begin{equation}
\lambda_\alpha/N(0)=Cg_\alpha^2/\hbar\omega_\alpha.
\label{boh}
\end{equation}
In the above expression, $N(0)$ is the density 
of levels at the Fermi energy, $\omega_\alpha$ are the phonon frequencies, and 
$C$ is $d_\alpha/2$, $d_\alpha$ being the 
degeneracy of the phonon state~\cite{notelast}. The results corresponding to the 
LUMO ($E_1^{\prime\prime}$)  
and the $A_1^\prime$ and $E_2^\prime$ phonons, are collected in Table I. 
We only report those phonons yielding a non--negligible partial coupling, 
i.e., $\lambda_\alpha/N(0) \geq 0.005$ meV.
For these phonons, we also show the value of the 
full width at half maximum (FWHM) of the Raman 
peaks ($\gamma_\alpha$), due to the decay to electron--hole pairs, widths 
which are connected 
to the partial couplings by the relation \cite{Gunnar}

\begin{equation}
 \gamma_\alpha = {2\pi \omega_\alpha^2 N(0) \lambda_\alpha \over d_\alpha}.  
\label{gamma}
\end{equation}

These widths show up as additional broadenings of the Raman peaks 
when the LUMO electronic
state becomes occupied, e.g., in the case of K$_4$C$_{70}$.
The differences between the widths of the Raman peaks of K$_4$C$_{70}$ and 
those of the peaks of bare
C$_{70}$ films have been measured~\cite{Dress2}, and are reported in the
last column of Table I. However, a direct comparison with the theoretical
results displayed in the fifth column of Table I 
is not straightforward for several reasons.
First of all, it must be observed that subtracting the 
phonon widths measured in pure C$_{70}$ 
from that measured in K$_{4}$C$_{70}$ 
leads, in some cases, to unrealistic negative values. 
Then, one should be aware of the fact that the frequencies and symmetry 
assignements of the experimental peaks 
are affected by some uncertainty. In any case, 
we followed the assignements given in
refs. \cite{Loosdr} and \cite{Benede}, which are different from those
of ref. \cite{Dress2}, since the former refer to low--temperature,
higher resolution spectra, and rely also on calculations of the 
phonon peaks strength~\cite{note_phon}. 
Finally, the value of $N(0)$ is not 
known with a high accuracy: we assume a value of
12 eV$^{-1}$ as suggested by Ref. \cite{Dress2}, but any value within the 
range 6--30 eV$^{-1}$ seems to be equally possible \cite{Gunnar}.
The total electron--phonon coupling constant $\lambda$ (sum of the partial 
coupling 
constants) computed making use of the results reported in the fifth column of
Table I, and of the empirical values of $\gamma_\alpha$ (seventh column of
Table I), are quite similar and of the order of 0.1. 
Keeping in mind the above--mentioned caveats, 
we may hence speak of an 
overall agreement between the results of the present calculation and the
experimental findings.
If $N(0)$ varies between $6$ and $30$ eV$^{-1}$, the theoretical total 
coupling constant varies between $0.05$ and $0.25$. These values  
must be compared with a value essentially equal to 1 obtained in  
the case of C$_{60}$ (cf., e.g.,~\cite{Gunnar} and \cite{BredaN}, as well as
references therein).

In a system where superconductivity is associated with the 
electron-phonon coupling, the transition temperature $T_c$ can be calculated
making use of McMillan's solution of the Eliashberg equation~\cite{McMill}
\begin{equation}
T_c={\hbar\omega_{\ln}\over 1.2 k_{\rm B}} \exp\left[-
    {1.04 (1+\lambda)\over \lambda-\mu^*(1+0.62\lambda)}
				\right], 
\label{Tc}
\end{equation}
where $\omega_{\ln}$ is a typical (logarithmic averaged) phonon frequency, 
and $\mu^*$ the Coulomb pseudopotential \cite{Schlut}.
While the above expression leads to $T_c\approx$ 10-15 K for C$_{60}$, making
use of typical values $\mu^*=$0.2-0.3, it predicts a vanishing value for
C$_{70}$, even allowing for uncertainties of the order of 50 to 100 $\%$ in
the estimated value of $\lambda$.

This result testifies to the fact that the transport properties in fullerites are
not simply dictated by the general features common to all these systems. 
Going from C$_{60}$ to C$_{70}$, deformation plays a very important role 
with respect to the electron-phonon interaction, by breaking the symmetries of 
electronic and vibrational levels, and by weakening their coupling. 

In fact, in order to better understand the origin of the large reduction
of the coupling strength in going from  C$_{60}$ to C$_{70}$, we have 
analyzed, for both clusters, the phonon eigenvectors of the low--frequency 
modes which couple to the electronic LUMO state, and compared them with the
spatial localization of the latter. 

In the low--frequency region, as illustrated in ref. \cite{LDA_phon}, it is
possible to recognize, with no ambiguity, the relation between 
a group of slightly splitted vibrational modes of C$_{70}$  and the 
corresponding quintuplet of an $H_{g}$ mode of C$_{60}$. This can be
done by classifying the modes of both clusters according to the largest
common symmetry subgroup, $C_{5v}$. We focus on the $H_{g}(2)$ mode of
C$_{60}$ at 408 cm$^{-1}$, which is associated with one of the largest 
$\lambda_{\nu}$. Following ref. \cite{LDA_phon},
we can trace the five modes associated with the $H_{g}(2)$ vibration
of  C$_{60}$  to the $A_{1}^{'}$,$E_{1}^{''}$ and $E_{2}^{'}$ modes of 
C$_{70}$ at 425, 415 and 442  cm$^{-1}$, respectively.
The $E_{1}^{''}$ mode does not couple with the LUMO of C$_{70}$,
hence about 2/5 of the coupling strength is lost. The remaining modes
are associated with a higher $\hbar \omega$ with respect to the case of
C$_{60}$, further reducing the coupling. Finally, a large reduction  effect
comes from the different spatial localization of the LUMO wave functions
in C$_{60}$ and C$_{70}$. In particular, we have explicitly verified that 
the shape of the molecular deformation associated with the above mentioned 
normal modes is quite similar in the two clusters~\cite{note_CPMD}. 
However, in the case of
C$_{60}$ the LUMO wave function is 
concentrated in the neighborhood of the
bonds which undergo the largest stretching while in  
C$_{70}$, the LUMO wave function is distributed on a larger region. 
We show in Fig. 1 the  
stereographic projection of C$_{60}$ LUMO wave function on a spherical surface 
with radius 3.4 $\AA$, while in Fig. 2 a similar projection - but on an 
ellipsoidal surface whose radii are 3.55 $\AA$ and 4.20 $\AA$ - is presented 
for the case of the LUMO wave function of C$_{70}$. 
In both figures, the thick continous lines mark 
the bonds which vary by more than 6\% under the phonon displacement 
fields. 

\section{Photoemission spectra}

The 
weakness of the electron-phonon matrix elements in the case of C$_{70}$ 
reflects itself also in the shape of the photoemission spectrum of
C$_{70}^-$, as is shown below. 
In the photoemission experiment, we are dealing with the process  
C$_{70}^-$$+ h\nu\to$ C$_{70}$ $+e^-$, where $h\nu$ is the incident photon 
and $e^-$ the emitted electron.

In order to define the ground state of C$_{70}^-$, we start from the full 
electron-phonon Hamiltonian, 
\begin{equation}
 H = \sum_i \varepsilon_i c^\dagger_i c_i + \sum_{\alpha}
 \hbar\omega_\alpha
 \Gamma^\dagger_{\alpha} \Gamma_{\alpha} +
 \sum_{i,j;\alpha} [V_{def}^{(\alpha)}]_{ij} c^\dagger_i c_j (
 \Gamma^\dagger_{\alpha} + \Gamma_{\alpha}),
\label{Heph}
\end{equation}
where the first term contains the Kohn-Sham energies $\varepsilon_i$, the
second term is the free phonon term and the last term is the usual bilinear 
electron-phonon coupling term. The 280 valence electrons of C$_{70}$ are 
assumed to be frozen (we call
$\psi_{\rm C}$ their wavefunction), while the extra electron
of C$_{70}^-$ is assumed to occupy one of the two (degenerate) lowest 
unoccupied states of
C$_{70}$ whose wavefunctions are labeled by $\psi_i$ ($i$=1,2). The
basis used for solving Eq.~(\ref{Heph}) is built up with these one electron
states coupled to 1, 2 or 3 different phonons (let us call $\phi_\alpha$ a phonon 
wavefunction). The matrix thus obtained, which turns out to be of 
the order of 5000 $\times$5000, is diagonalized by using the Lanczos method. 
We denote by $E_0$ its lowest 
eigenvalue and by $\Psi_0$ the corresponding wavefunction. This 
should then describe the ground state of the charged molecule C$_{70}^-$,
\begin{equation}
\Psi_0 = \sum_i a_i~\psi_i~\psi_{\rm C} +
	 \sum_{i,\alpha}~b_i(\alpha)~\psi_i~\psi_{\rm C}~\phi_\alpha +
	 \sum_{i,\alpha,\alpha^\prime}~c_i(\alpha,\alpha^\prime)~\psi_i~\psi_{\rm
	 C}~\phi_\alpha~\phi_{\alpha^\prime} +...
\label{totpsi}
\end{equation}

Assuming that the emitted electron does not 
interact with the system left behind (sudden approximation~\cite{Hed69}), 
and that its wavefunction is described by a plane wave, the transition 
probability for the emission process,
from the initial state $\Psi_0$ to a final state of energy $E_f$ in which the
neutral C$_{70}$ has 1, 2 or 3 phonons $\alpha$ excited (we denote by
$\Phi_f$ the corresponding 
total vibrational wavefunction), is given by the Fermi 
golden rule,
\begin{equation}
 W  \sim \sum_f  \vert\langle e^{-i{\vec k}\cdot{\vec r}} \psi_{\rm C}~
		 \Phi_f\vert V_{\rm ext}\vert\Psi_0\rangle\vert^2~
		 \delta(h\nu+E_0-E_f-\varepsilon),
\label{golden}
\end{equation}
where $V_{\rm ext}$ is the (dipole) external field, $E_f=\sum_\alpha
\hbar\omega_\alpha$ (the energy of the 280 frozen electrons does not appear
neither in the initial nor in the final state), and $\varepsilon$ is 
the electron energy. If we substitute Eq.~(\ref{totpsi}) into 
Eq.~(\ref{golden}), 
and assume that the dominant terms are those corresponding to the 
LUMO (i.e., to the state $\psi_1$), we can factor out the dipole matrix 
element, and we are left with
\begin{eqnarray}
 W_0 \sim \vert a_1\vert^2~\delta(h\nu+E_0-\varepsilon) &+&
 \sum_\alpha \vert b_1(\alpha)\vert^2~\delta(h\nu+E_0-\hbar\omega_\alpha-\varepsilon)
 \nonumber \\
 &+&\sum_{\alpha,\alpha^\prime} \vert c_1(\alpha,\alpha^\prime)\vert^2~
 \delta(h\nu+E_0-\hbar\omega_\alpha-\hbar\omega_{\alpha^\prime}-\varepsilon).
\label{spectr}
\end{eqnarray}

We have plotted the results for $W_0$ in Fig.~3. The scale has been set in 
such a way that $h\nu+E_0\equiv 0$ and the numbers in the horizontal axis
correspond to $\varepsilon$. While the shoulder at 300 cm$^{-1}$ and the 
tail extending from 400 to 800 cm$^{-1}$ indicate the presence of satellite 
peaks containing 1 and 2 phonons (as emphasized by the thin curves), the 
rather smooth behavior of the cross section and the strength of the main
peak with respect to the satellites, as compared to the
corresponding quantities in the case of C$_{60}$, 
testify to the fact that the electron-phonon coupling in C$_{70}^-$ is 
much weaker than in C$_{60}^-$  (see inset of Fig.~3). 
While the overall
behavior of the photoemission cross section predicted by theory is 
confirmed
by the experimental data, lack of resolution does not allow for a detailed
comparison.  

\section{Conclusions}

The calculated electron-phonon coupling in C$_{70}$ is 
found to be about
one order of magnitude weaker than in C$_{60}$.
In particular, we estimate the total dimensionless electron--phonon coupling 
constant $\lambda$ to be about $0.1$. 
The decrease with respect to C$_{60}$ is understood
on the  basis of the symmetry reduction, and the different shape and
spatial localization of the LUMO wavefunction in C$_{70}$.  
The  calculated $\lambda$ value is
consistent with the lack of the superconducting phase transition reported
in the literature at temperatures as low as 1 K. This value
agrees also with the overall features of two relevant experimental 
measurements: 
the broadening of Raman peaks in K$_4$C$_{70}$ and the
photoemission cross section in C$_{70}^-$. In the latter case, 
experiments with higher resolution would be required to test in detail 
the theoretical predictions.

\section*{Acknowledgments}

This work has been partially supported by the INFM advanced research program 
CLASS.

\begin{table} 
\caption{Reduced electron--phonon matrix element $g_\alpha$ and partial coupling 
constants $\lambda_\alpha/N(0)$ for the LUMO electron state and selected phonons 
of C$_{70}$. The first two columns report the calculated 
phonon
frequencies and symmetries obtained within a bond-charge model
calculation~{\protect \cite{BCM}}.  
In the fifth column we also show the calculated widths of the phonon peaks (Eq.
({\protect \ref{gamma}})).
For comparison, the last two columns give the experimental frequencies and the mesured 
widening $\gamma$ of the Raman peaks in going from pure C$_{70}$ to 
K$_{4}$C$_{70}$ films.}
\vspace{0.3cm}
\begin{tabular}{ccccccc}
\multicolumn{5}{c}{Theory}&\multicolumn{2}{c}{Experiment \protect{\cite{Dress2}}}\\ 
Frequency [cm$^{-1}$]  & symm. & $g_\alpha$ [meV] 
& $\lambda_\alpha/N(0)$ [meV] & $\gamma$ [cm$^{-1}$] & Frequency [cm$^{-1}$]
& $\gamma$ [cm$^{-1}$] \\ \hline
230 & $E_2^{'}$ & 4.77 & 0.798 & 2.36 & 226 & 10.7 \\
271 & $A_1^{'}$ & 21.9 & 7.136 & 58.7 & 256 & -1.5 \\
308 & $E_2^{'}$ & 6.03 & 0.951 & 5.06 & --- & --- \\
425 & $A_1^{'}$ & 3.59 & 0.122 & 2.48 & 394 & 6.3 \\
442 & $E_2^{'}$ & 3.76 & 0.258 & 2.82 & 408 & -0.5 \\
448 & $A_1^{'}$ & 4.95 & 0.220 & 4.96 & 454 & 4.9 \\
508 & $E_2^{'}$ & 3.26 & 0.169 & 2.44 & --- & --- \\
590 & $E_2^{'}$ & 3.09 & 0.130 & 2.55 & --- & --- \\
625 & $A_1^{'}$ & 1.95 & 0.025 & 1.08 & 567 & 3.5 \\
701 & $A_1^{'}$ & 1.71 & 0.017 & 0.93 & 702 & 13.9 \\
736 & $E_2^{'}$ & 6.98 & 0.533 & 16.2 & 712 & 14.3 \\
764 & $E_2^{'}$ & 2.48 & 0.065 & 2.12 & 737 & 0.5 \\
1089 & $E_2^{'}$ & 2.66 & 0.052 & 3.48 & --- & --- \\
1200 & $A_1^{'}$ & 1.73 & 0.010 & 1.62 & 1181 & 12.9 \\
1223 & $A_1^{'}$ & 1.46 & 0.007 & 1.18 & 1228 & 6.5 \\
1368 & $A_1^{'}$ & 6.45 & 0.123 & 25.7 & 1444 & 9.0 \\
1459 & $A_1^{'}$ & 6.79 & 0.127 & 30.4 & 1459 & --- \\
\end{tabular}

\end{table}

\begin{figure}
Fig. 1: Stereographic projection on a plane of the values taken by the 
squared C$_{60}$ LUMO wavefunction on a
spherical surface of radius $3.4\ \AA$. The thick lines mark the bonds 
that vary more than 6\% by effect of the displacement field associated with 
the $H_g$ phonon at 408 cm$^{-1}$. 
\end{figure}

\begin{figure}
Fig. 2: Stereographic projection on a plane of the values taken by the 
squared C$_{70}$ LUMO wavefunction on an
ellipsoidal surface of radii $3.55\ \AA$ and $4.20\ \AA$. 
The thick lines mark the bonds 
that vary more than 6\% by effect of the displacement field associated with 
the $A_1^\prime$ phonon at 425 cm$^{-1}$.
\end{figure}

\begin{figure}
Fig. 3: Calculated photoemission spectrum for C$_{70}^-$ (solid line). In 
the inset, the photoemission spectrum for C$_{60}^-$ from reference 
{\protect\onlinecite{BredaN}} is shown.
The contributions coming from zero phonons (long--dashed line), 
one phonon (dash--dotted line) and two phonons (dashed line) are also plotted 
separately. See the text for a discussion.  
A broadening with Gaussians of 40 meV width has been used in constructing the 
photoemission spectra.
\end{figure} 


\begin{thebibliography}{9}
\bibitem{BCM} We used the model of ref. \cite{Benede} with the following  
              set of parameters (original values in parenthesis): 
	      $\frac{Y^{2}}{\epsilon} = 1.63\ (1.628)$, 
	      $\rho = 0.43\ (0.431)\ \AA$, $c = 0.36\ (0.511)\ \AA^{-1}$, 
	      $\beta = 48.7\ (50.5) \cdot 10^4$ dyne/cm and 
	      $\Phi_{0} = 28.86\ (28.86) \cdot10^{-12} \mbox{erg}$. The slight  
	      differences with respect to the original values are due 
	      to the use of an improved set of equilibrium coordinates. The 
              change in the value of $c$ has a little effect on the BCM 
              calculation, as can be understood from its definition in~\cite{Benede}.
\bibitem{notelast} As should be evident from the above discussion, 
                   $g_{\alpha}$ includes the factor 
                   $\sqrt{\hbar \over 2 m \omega}$ coming from
                   the matrix element of $\vec Q_{N}$ between 
                   zero-- and one--phonon vibrational states (in the literaure,
                   other definitions of $g_\alpha$ are sometimes found).
\bibitem{note_phon} As an example, let us consider the low--energy
                  peak measured at 256 cm$^{-1}$ in ref. 
                  {\protect\cite{Dress2}}, which in refs. 
                  {\protect\cite{Loosdr}} and {\protect\cite{Benede}} 
                  is resolved into two distinct modes.
\bibitem{note_CPMD} We have also verified that this conclusion does not depend 
                    on the model chosen to calculate the phonons: for the modes 
                    at hand, the displacement fields extracted from 
                    Car-Parrinello molecular dynamics 
                    {\protect \cite{LDA_phon}} coincide with those of 
                    BCM within a few percent.
\end{thebibliography}
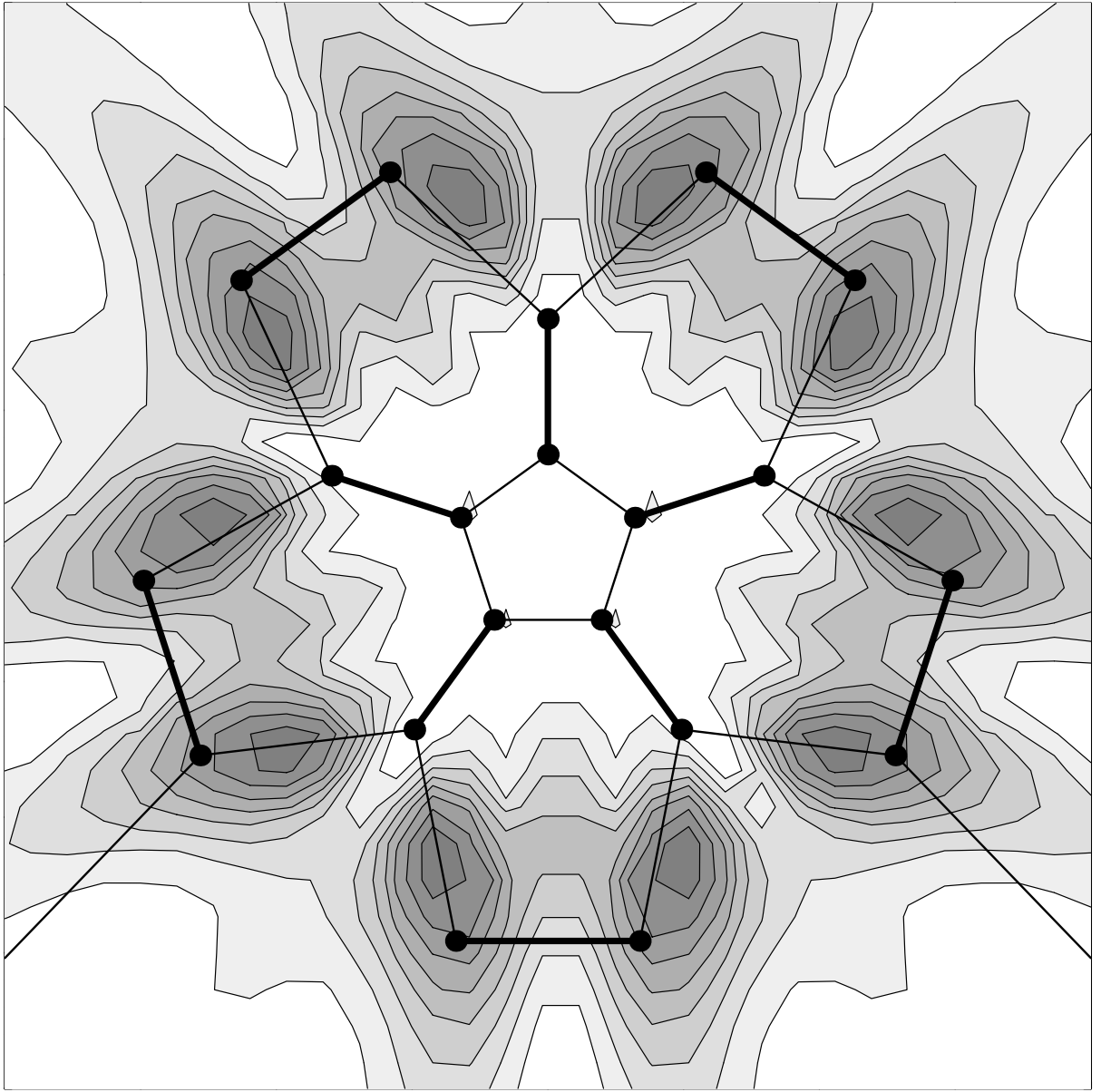
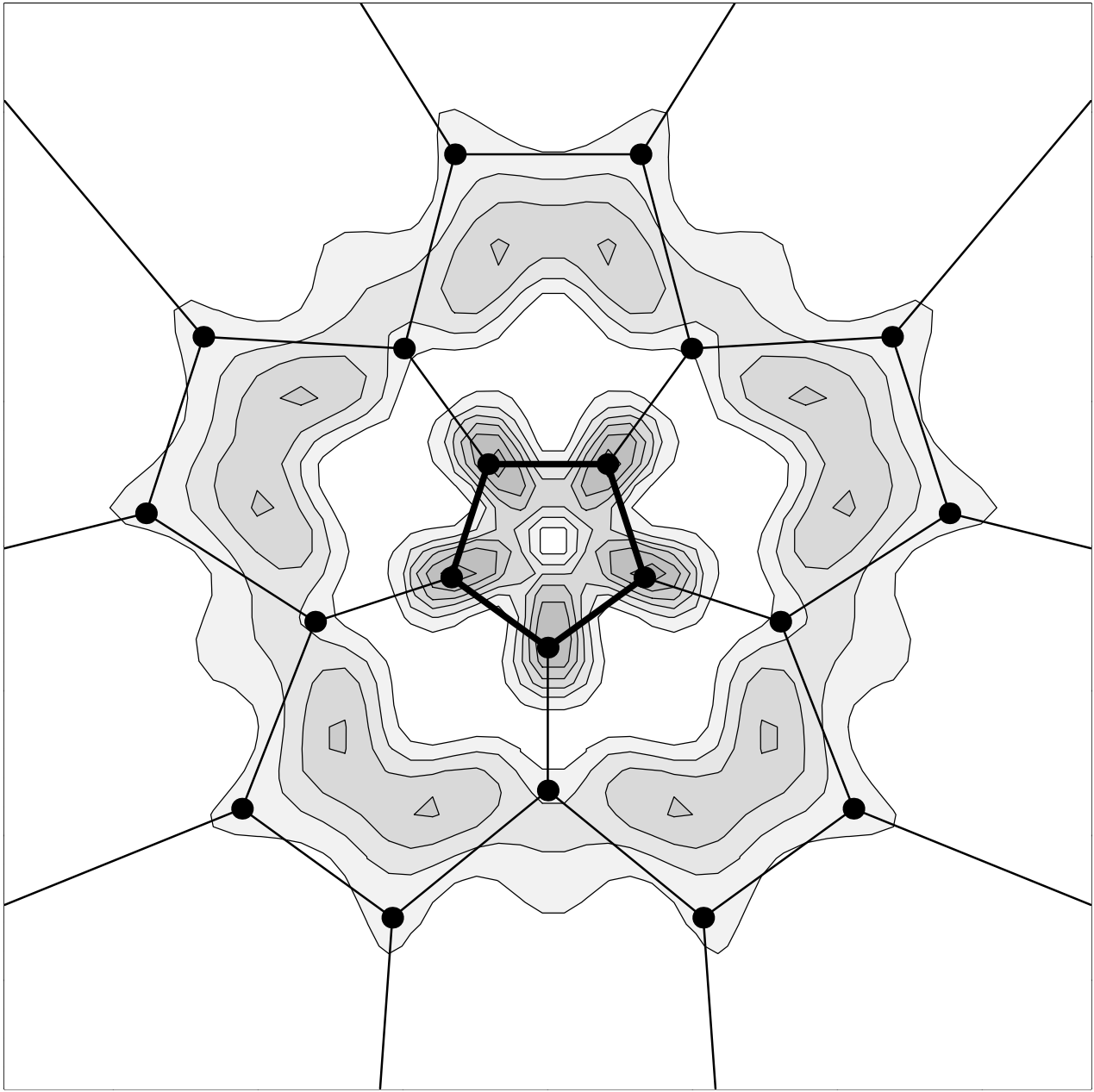
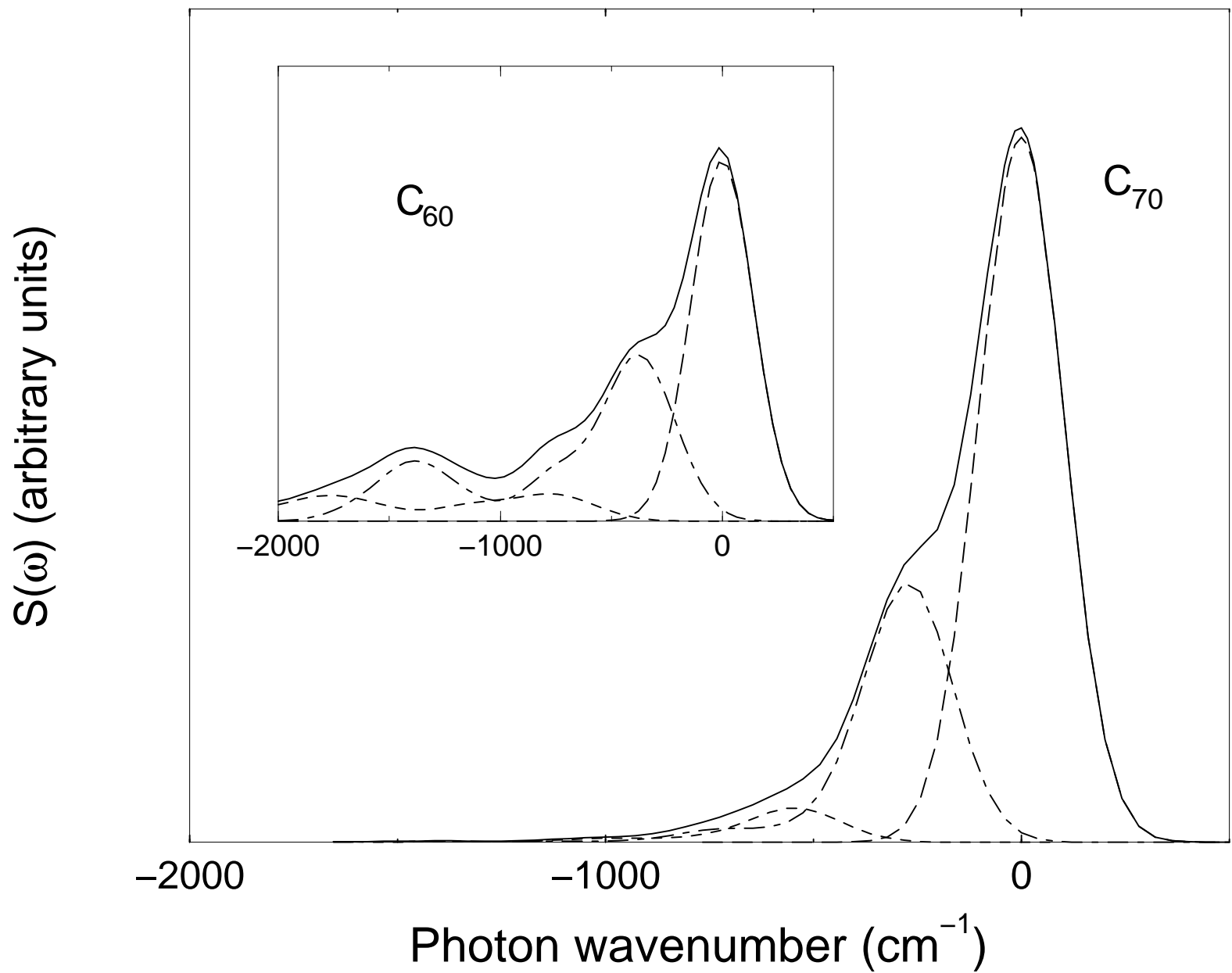
\end{document}